\documentclass[runningheads]{llncs}
\usepackage[utf8]{inputenc}
\usepackage{graphicx}
\usepackage{hyperref}
\usepackage{amsmath}
\usepackage{hyperref}
\usepackage{subfig}
\usepackage{caption}
\usepackage[table]{xcolor}
\usepackage{float}
\usepackage{paralist}
\usepackage{url}
\usepackage{tabto}
\usepackage{latexsym}
\usepackage{textcomp}
\usepackage{amsfonts}
\usepackage{hyphenat}
\usepackage{xcolor}
\usepackage{amssymb}
\usepackage{algorithm}
\usepackage[noend]{algpseudocode}
\usepackage{listings}
\definecolor{powderBlue}{RGB}{147, 207, 207}
\definecolor{blueGreen}{RGB}{89, 189, 192}
\definecolor{metallicSeaweed}{RGB}{6, 134, 146}
\definecolor{viridianGreen}{RGB}{1, 146, 150}
\definecolor{burntSienna}{RGB}{220, 120, 85}
\definecolor{melon}{RGB}{246, 183, 161}
\definecolor{rust}{RGB}{190, 51, 2}
\definecolor{grey}{RGB}{183, 183, 183}

\begin{document}
\title{FunMap: Efficient Execution of Functional Mappings for Knowledge Graph Creation}
\titlerunning{FunMap: Efficient Execution of Functional Mappings for KGC}
\author{Samaneh Jozashoori\inst{1}\orcidID{0000-0003-1702-8707}, 
        David Chaves-Fraga\inst{2}\orcidID{0000-0003-3236-2789},
        Enrique Iglesias\inst{1,3}\orcidID{0000-0002-8734-3123},\\
        Maria-Esther Vidal\inst{1}\orcidID{0000-0003-1160-8727},Oscar Corcho\inst{2}\orcidID{0000-0002-9260-0753}
        }
%
\authorrunning{Jozashoori et al.}
%
\institute{TIB Leibniz Information Center for Science and Technology \& L3S, Germany\\
\email{\{samaneh.jozashoori,maria.vidal\}@tib.eu}
\and
Ontology Engineering Group, Universidad Polit{\'e}cnica de Madrid, Spain\\
\email{\{dchaves,ocorcho\}@fi.upm.es}
\and
University of Bonn\\
\email{s6enigle@uni-bonn.de}
}
\maketitle              
\begin{abstract}
Data has exponentially grown in the last years, and knowledge graphs constitute powerful formalisms to integrate a myriad of existing data sources. 
 Transformation functions -- specified with function-based mapping languages like FunUL and RML+FnO -- can be applied to overcome interoperability issues across heterogeneous data sources. However, the absence of engines to efficiently execute these mapping languages hinders their global adoption. We propose FunMap, an interpreter of function-based mapping languages; it relies on a set of lossless rewriting rules to push down and materialize the execution of functions in initial steps of knowledge graph creation. Although applicable to any function-based mapping language that supports joins between mapping rules, FunMap feasibility is shown on RML+FnO. FunMap reduces data redundancy, e.g., duplicates and unused attributes, and converts RML+FnO mappings into a set of equivalent rules executable on RML-compliant engines. We evaluate FunMap performance over real-world testbeds from the biomedical domain. The results indicate that FunMap reduces the execution time of RML-compliant engines by up to a factor of 18, furnishing, thus, a scalable solution for knowledge graph creation.

\end{abstract}

\section{Introduction}
\label{sec:intro}
Knowledge graphs (KGs) have gained momentum due to the explosion of available data and the demand for expressive formalisms to integrate factual knowledge spread across various data sources~\cite{abs-2003-02320}.
KG creation requires the description of schema alignments among data sources and an ontology, as well as the specification of methods to curate and transform data collected from the input sources into a unified format. A rich spectrum of mapping languages has been proposed to specify schema-ontology alignments across data sources implemented in a variety of semi-structured and structured formats; exemplar approaches include R2RML~\cite{das2012r2rml}, RML~\cite{dimou2014rml}, and xR2RML~\cite{MichelDFM15}. Furthermore, function-based mapping languages \cite{de2017declarative,debruyne2016r2rml,junior2016funul,vu2019d} are equipped with abstractions that enable interoperable and reusable specifications of data transformations by means of user-defined functions. Moreover, formalisms like RML+FnO~\cite{de2017declarative} combine the Function ontology and RML, enabling declarative specification of the schema-ontology alignments and data transformations that define the process of KG creation. Albeit expressive, existing mapping languages lack efficient interpreters able to scale up to complex KG creation scenarios. The incoming data avalanche urges KG creation approaches capable of integrating large and diverse data, and efficiently transforming this data to comply with application-specific KG formats.   

\noindent \textbf{Problem and Objectives:} We tackle the problem of scaled-up KG creation from functional mapping rules and study the impact of functions when applied to large data sources with a high data duplication rate. A KG creation process is defined as a data integration system~\cite{lenzerini2002data}. Mappings among data sources and the system ontology are expressed using the RDF mapping language (RML) \cite{de2017declarative} and the Function Ontology (FnO); they define how the ontology concepts are populated with data from the sources in the resulting KG. We aim at transforming complex data integration systems composed of large data sources and mappings with functions into equivalent ones that generates the same KG but in less time.

\noindent \textbf{Our Proposed Approach:} We present FunMap, an interpreter of RML+FnO, that converts a data integration system defined using RML+FnO into an equivalent one where RML mappings are function-free. FunMap resembles existing mapping translation proposals (e.g., \cite{AliW19,corcho2020towards,junior2016funul}) and empowers a KG creation process with optimization techniques to reduce execution time. Transformations of data sources include the projection of the attributes used in the RML+FnO mappings. They are supported on well-known properties of the relational algebra, e.g., the pushing down of projections and selections into the data sources, and enable not only the reduction of the size of data sources but also the elimination of duplicates. Additionally, FunMap materializes functions --expressed in FnO-- and represents the results as data sources of the generated data integration system; the translation of RML+FnO into RML mappings that integrate the materialization of functions is performed using joins between the generated RML mappings. The combination of data source and function transformations results in data integration systems where only the data required to execute the RML mappings are retained. The computation of the functions used in the original data integration system is performed once. As a result, the new data integration system's execution is sped up while the same knowledge graph is generated. 

\noindent\textbf{Contributions.}
\begin{inparaenum}[\bf i\upshape)]
    \item FunMap, an interpreter of RML+FnO that resorts to syntax-based translation~\cite{aho1986compilers} to push down projections and selections, and materialize functions. 
    \item Empirical evaluations of the performance of FunMap in real-world testbeds with data of various formats (CSV and Relational), sizes, and degrees of duplication that show reductions in KG creation time by up to a factor of 18.
\end{inparaenum}


The remainder of this paper is structured as follows: Section \ref{sec:example} motivates our work using a use case from the biomedical domain. Section \ref{sec:approach} describes the set of rewriting and optimization rules that assemble FunMap, while the experimental results are reported in Section \ref{sec:eval}. Finally, Section \ref{sec:relatedwork} presents the related work and Section \ref{sec:conclusions} outlines the main conclusions of the paper and future lines of work. 

\section{Preliminaries and Motivating Example}
\label{sec:example}
\begin{figure}[t!]
\centering
\includegraphics[width=0.9\textwidth]{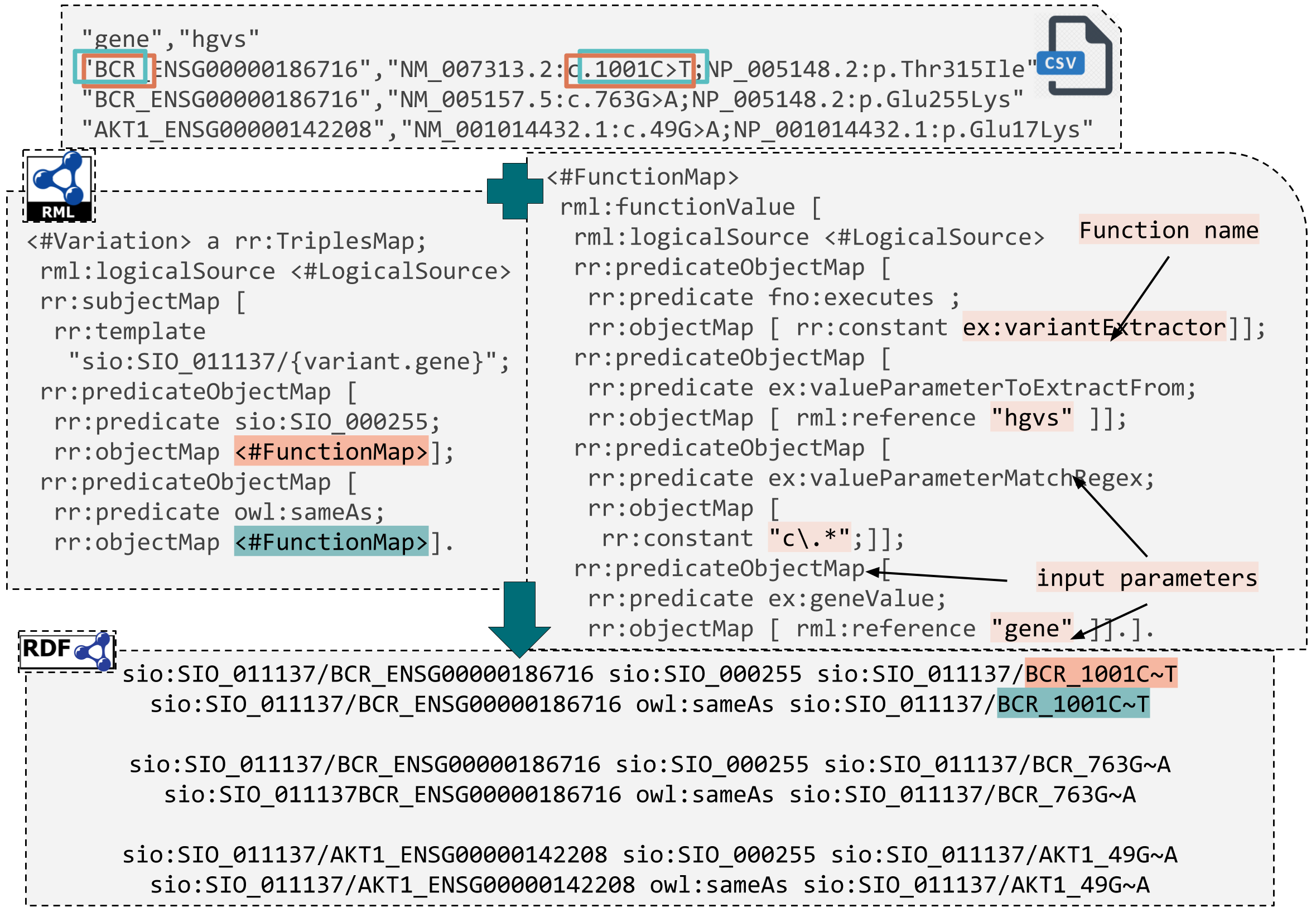}
\caption{\textbf{Motivating example.} Knowledge graph construction using RML+FnO mapping rules for the biomedical domain. The input source in the top is transformed to RDF output (at the bottom) through the processing of the mapping (middle) where the transformation functions are defined. Repeated computations of a function negatively impacts on the performance of an RML engine.} 
\label{fig:motivatingExample}
\end{figure}
\subsection{Preliminaries}
\label{sec:preliminaries}
The RDF Mapping Language (RML) extends the W3C-standard mapping language R2RML with logical sources (a.k.a. \verb|logicalSource|) in heterogeneous formats (e.g., CSV, Relational, JSON, and XML). As the W3C-standard R2RML, \verb|TriplesMap| corresponds to mapping rules where the resources (a.k.a. \verb|subjectM|-\verb|ap|) of an RDF class and their properties (a.k.a. \verb|predicateMap|) are assigned to values (a.k.a. \verb|objectMap|) based on logical data sources. An \verb|objectMap| can be also defined as a reference or a join with the \verb|subjectMap| in another \verb|TriplesMap| (a.k.a. \verb|RefObjectMap| and \verb|joinCondition|, respectively). \verb|subjectMap|, \verb|predica|-\verb|teMap|, and \verb|objectMap| are also referred as \verb|TermMap| in general; they generate RDF terms. FnO is an ontology for describing transformation functions declaratively; FnO and RML relationship is described in \cite{de2017declarative}. Accordingly, the \verb|FunctionMap| class is introduced in RML; it defines transformation functions in any part of the \verb|TriplesMap| (\verb|subjectMap|, \verb|predicateMap|, or \verb|objectMap|). These concepts are illustrated in the next example and highlighted in \autoref{fig:motivatingExample}.

\subsection{A Real-World Example from the Biomedical Domain}
Our work is motivated by the challenges revealed during genomic variant reconciliation while creating a biomedical knowledge graph. 
Although the vast majority of the single variations in the genome of a person causes no disease, benign variants can appear in sequenced genomic data repeatedly. In addition to the large heterogeneous volumes generated during genome sequencing and analysis, high-frequency of genomic variants impose data integration challenges while collecting genomic data from different sources. 
Additionally, genomic variants are expressed in diverse standard formats~\cite{den2016hgvs} and reported at DNA, RNA, or protein level. Moreover, this representation can be done according to any of the accepted terminologies and genomic reference versions. Unified representations for variants are required
to semantically recognize and integrate equivalent variants residing in different data sources. Variant representations can result from a composition of several factors, such as gene name, genomic position, and residue alteration. 
Pre-processing functions (e.g., FnO functions) are needed to extract and compose values from different attributes from each data source and generate such a combined representation of variants.
These functions are part of the data integration system's mapping rules that define the KG creation process. 

\autoref{fig:motivatingExample} depicts a mapping rule in  RML+FnO where the \verb|FunctionMap| class is utilized. Consider that according to the \verb|LogicalSource| provided in this example, a \verb|FunctionMap| is defined in the mapping rules to create a unified representation for a variant by extracting the values of ``gene name'' (e.g., BCR) from the attribute \verb|gene| and 
``coding alteration'' (e.g., c.1001C\textgreater T)  from the attribute named \verb|hgvs| and combine them (e.g., BCR\textunderscore1001C\textasciitilde T). Current approaches evaluate \verb|FunctionMap| for each \verb|variant|, which can be expensive in presence of large data sources. Nevertheless, the large number of redundant values leaves room for the scalable transformations to execute functional mappings. 

\section{The FunMap Approach}
\label{sec:approach}
\begin{table}[h!]
\normalsize
\centering
\caption{Summary of the notation used for defining FunMap}
\label{tab:notations}
\resizebox{1.0\textwidth}{!}{%
\begin{tabular}{|l|l|}
\hline
\textbf{Notation} & \textbf{Explanation} \\ \hline
$DIS_G=\langle O,S,M \rangle$ & Data Integration System which creates a KG $G$ \\ \hline
$O$ & Unified Ontology of $DIS_G=\langle O,S,M \rangle$\\ \hline
$S$ & Finite set of Data Sources $S_i$ of $DIS_G=\langle O,S,M \rangle$\\ \hline
$M$ & Finite set of \texttt{TriplesMap}s $T_i$ in $DIS_G=\langle O,S,M \rangle$\\ \hline
$\emph{RDFize}(.)$ & A function producing RDF triples from a data integration system\\ \hline
$T'_i$ and $T'_k$ & \texttt{TriplesMap}s resulting of applying MTRs \\ \hline
$F_i$ & A Transformation Function in a \texttt{TriplesMap} in $M$ \\ \hline
$S'$ & Finite set of Data Sources $S'_i$ resulting of applying DTRs\\ \hline
$M'$ & Finite set of Mapping Rules $M'_i$ resulting of applying MTRs\\ \hline
$S_i^{output}$ & Data source resulting of applying DTR1, with attributes $o'_i$ and $a'_i$ \\ 
& representing the materialization of a transformation function $F_i$ \\ \hline
$S_i^{project}$ &  Data source resulting of applying DTR2 \\ \hline
\end{tabular}%
}
\end{table}
FunMap is an interpreter of data integration systems $DIS_G = \langle O,S,M \rangle$, where $O$ stands for a unified ontology, and $S$ and $M$ represent sets of sources and mapping rules, respectively \cite{lenzerini2002data}. The evaluation of $DIS_G$ (a.k.a. $\emph{RDFize}(DIS_G)$) results into a knowledge graph $G$ that integrates data from $S$ according to the mapping rules in $M$; entities and properties in $G$ are described in terms of $O$. A complex data integration system $DIS_G$ consists of large data sources with high-duplicated data and mapping rules including functions for both schema-ontology alignments and data transformations. FunMap converts $DIS_G$ into an equivalent data integration system that creates the same knowledge graph but in less time. \autoref{tab:notations} summarizes the notation utilized in the FunMap approach.

\noindent \textbf{Problem Statement:} 
Given a data integration system $DIS_G$=$\langle O,S,M\rangle$, the problem of scaled-up knowledge graph creation from functional mappings requires the generation of a data integration system 
$DIS'_G$=$\langle O,S',M' \rangle$:
\begin{itemize}
\renewcommand{\labelitemi}{$\bullet$}
\item The knowledge graphs resulting of the evaluations of both data integration systems are the same, i.e., $\emph{RDFize}(DIS'_G = \langle O,S',M' \rangle)$=$\emph{RDFize}(DIS_G = \langle O,S,M \rangle)$ where $RDFize(.)$ is a function producing RDF triples utilizing the input data integration system.
\item The execution time of $\emph{RDFize}(DIS'_G=\langle O,S',M' \rangle)$ is \emph{less than} the execution time of $\emph{RDFize}(DIS_G=\langle O,S,M \rangle)$.
\end{itemize}

\begin{figure}[t!]
\includegraphics[width=\textwidth]{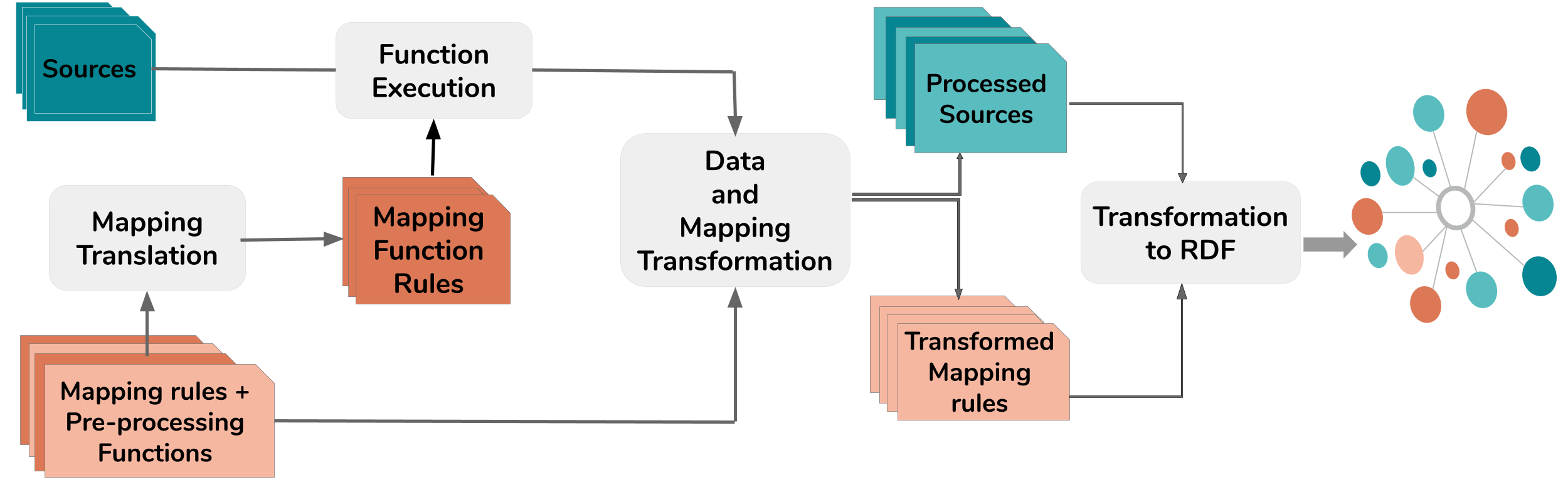}
\caption{\textbf{The FunMap approach}} 
\label{approach}
\end{figure}

\noindent \textbf{Solution:} FunMap implements a heuristic-based approach; it relies on the assumption that eliminating duplicates, maintaining in the data sources only the attributes mentioned in the mappings, and materializing the functions in the mappings, reduces the execution time of knowledge graph creation process. 
FunMap receives a data integration system $DIS_G$=$\langle O,S,M \rangle$ where the mappings $M$ are expressed in RML+FnO. FunMap interprets the mappings in $M$ and converts $DIS$ into the data integration system $DIS'_G$ in which the mappings $M'$ are function free and duplicates in the data sources $S'$ are reduced. \autoref{approach} depicts the FunMap approach; it performs a syntax-based translation of the mappings in $M$ and ensures that each redundant function is evaluated exactly once on the same data values. 
FunMap transforms $S$ to $S'$ by means of data transformation rules (DTR1 and DTR2). For each $F_i$ over a given $S_i$, DTR1 creates a temporal source $S'_i$ that includes the attributes from $S_i$ that correspond to the input of $F_i$; it also generates a source $S_i^{output}$ that contains the attributes in $S'_i$ and attributes representing the output of $F_i$. For each \verb|FunctionMap| defined over a source $S_i$, DTR2 creates a source $S_i^{project}$ that includes all attributes of $S_i$ used in the \verb|FunctionMap|. Additionally, FunMap converts mapping rules that include functions by using mapping transformation rules (MTRs); a \verb|FunctionMap| is transformed into \verb|FunctionMap|s without functions while connected by \verb|joinCondition|s; initially, $S'$ and $S$ are equal, as well as $M'$ and $M$. Properties \ref{property:p1}, \ref{property:p2}, and \ref{property:p3} state the pre- and post-conditions of DTRs and MTRs.

\subsection{Transformation Rules in FunMap}
\label{subsec:transformation}
\begin{figure}[b!]
\includegraphics[width=\textwidth]{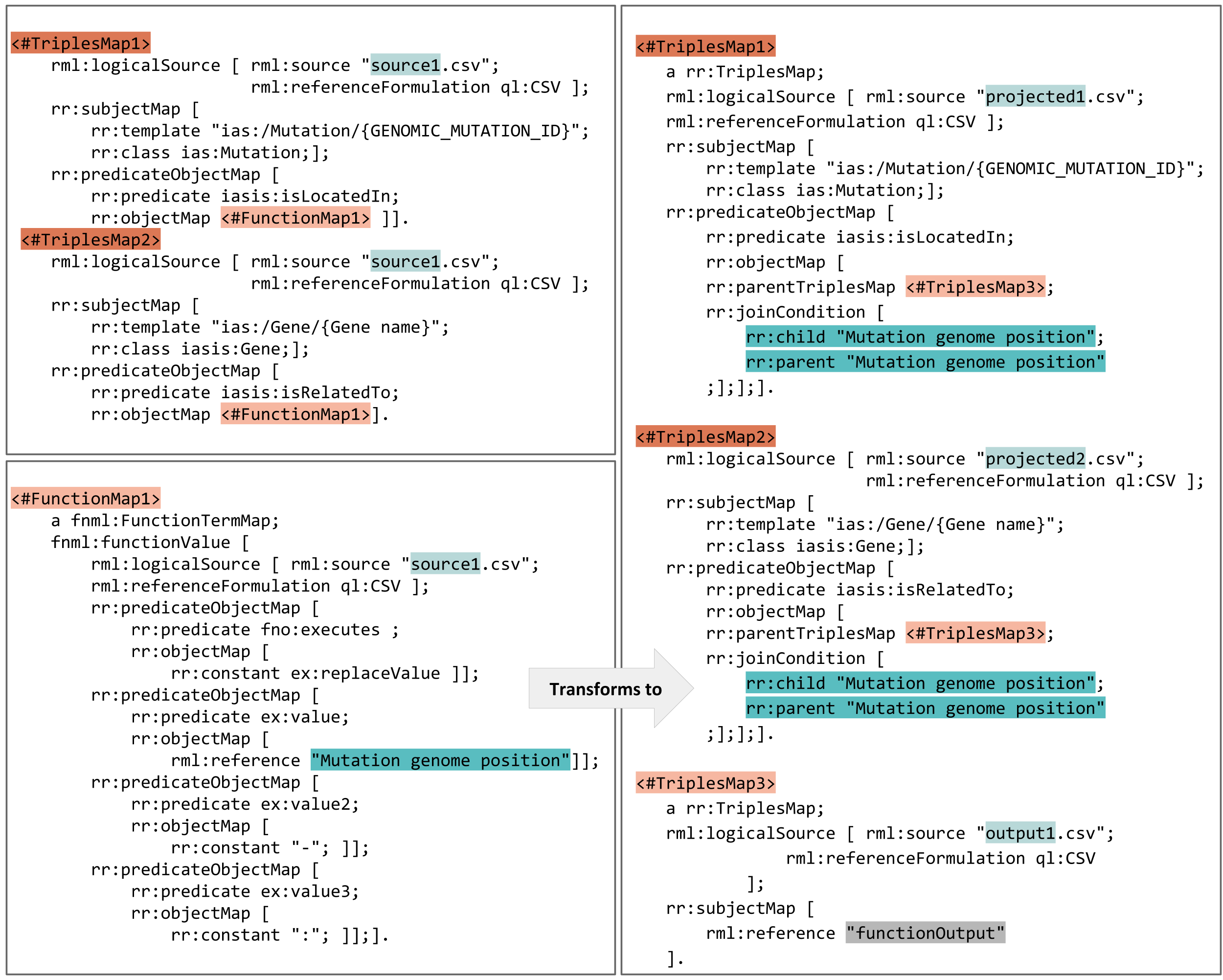}
\caption{\textbf{Example of DTR and Object-based MTR.} On the left, an exemplary mapping including two TriplesMaps and a FunctionMap provided by the original data integration system. On the right side, the mappings are transformed by FunMap including two new TriplesMaps and one new TriplesMap.} 
\label{fig:objectBased}
\end{figure}
The FunMap syntax-based translation component parses \verb|FunctionMap|s exactly once, i.e., \verb|FunctionMap|s repeated in various mappings are not evaluated more than once. Given \verb|FunctionMap|s, original data sources, and mappings, FunMap executes transformation rules on data sources and mappings, accordingly. Meanwhile, given the transformed data sources, FunMap detects that a \verb|FunctionMap| has been computed for a given value and avoids repeating this computation. As an outcome, FunMap provides 
\begin{inparaenum}[\bf a\upshape)]
    \item a new set of data sources $S'$ consisting of the original ones in conjunction with transformed data sources,  and 
    \item a set $M'$ of transformed function-free mappings. 
\end{inparaenum}    
FunMap is loyal to the formats of data sources and mappings. Thus, any RDF mapping language is compatible with the process implemented in FunMap, as far as the language enables the definition of joins between mapping rules. Next, we present the transformation rules. 

\noindent \textbf{Data Source Transformation Rules (DTRs):}
Considering the fact that a \verb|TriplesMap| may only be used some attributes of a dataset, FunMap relies on the properties of the relational algebra and performs DTRs to project only the attributes mentioned in the \verb|TriplesMap|. DTRs are followed by transformation rules (MTRs) that update mappings defined over the transformed data sources. 

\noindent \textbf{DTR1: Projection of Functional Attributes:}
For each transformation function $F_i$ over a given source $S_i$ in the set of data sources $S$, FunMap projects all attributes $a'_i$ in $S_i$ that are input attributes of $F_i$, into a temporal data source $S'_i$ followed by duplicate removal. Subsequently, it evaluates $F_i$ over $S'_i$ and stores the results into the attribute $o_i$. Lastly, it creates a new data source $S_{i}^{output}$ with the attributes $a'_i$ and $o_i$; $S_{i}^{output}$ is added to $S'$. 
\begin{figure}[t!]
\includegraphics[width=\textwidth]{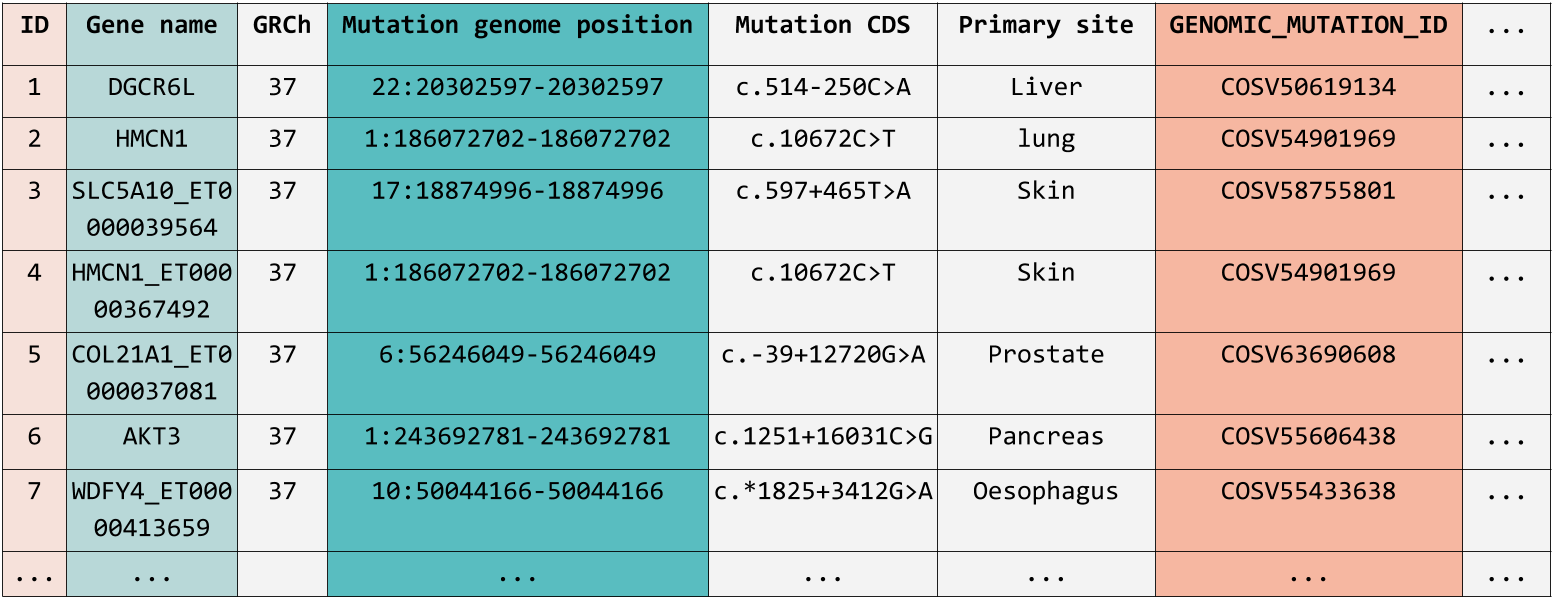}
\caption{\textbf{Original data source for KG creation.} The data source includes many attributes among which only a few are required by the transformation function or function-free mappings in the process of knowledge graph creation.} 
\label{fig:OriginalDatasource}
\end{figure}
\begin{figure}[t!]
 \centering
 \subfloat[Projected1]{
      \includegraphics[width=0.32\columnwidth]{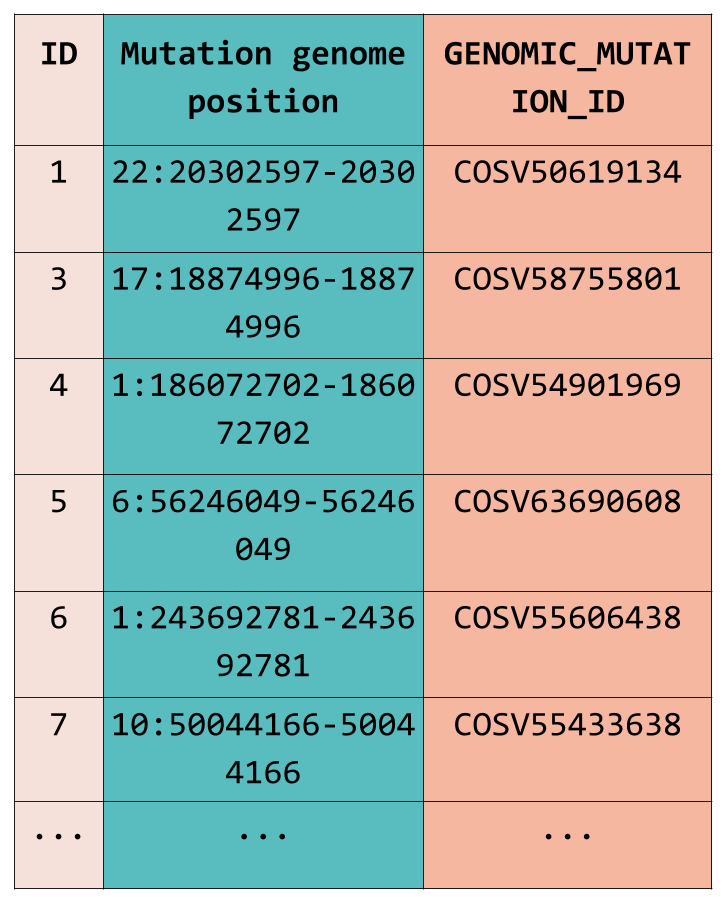}
    \label{fig:projected1}}
  \subfloat[Projected2]{
\includegraphics[width=0.32\columnwidth]{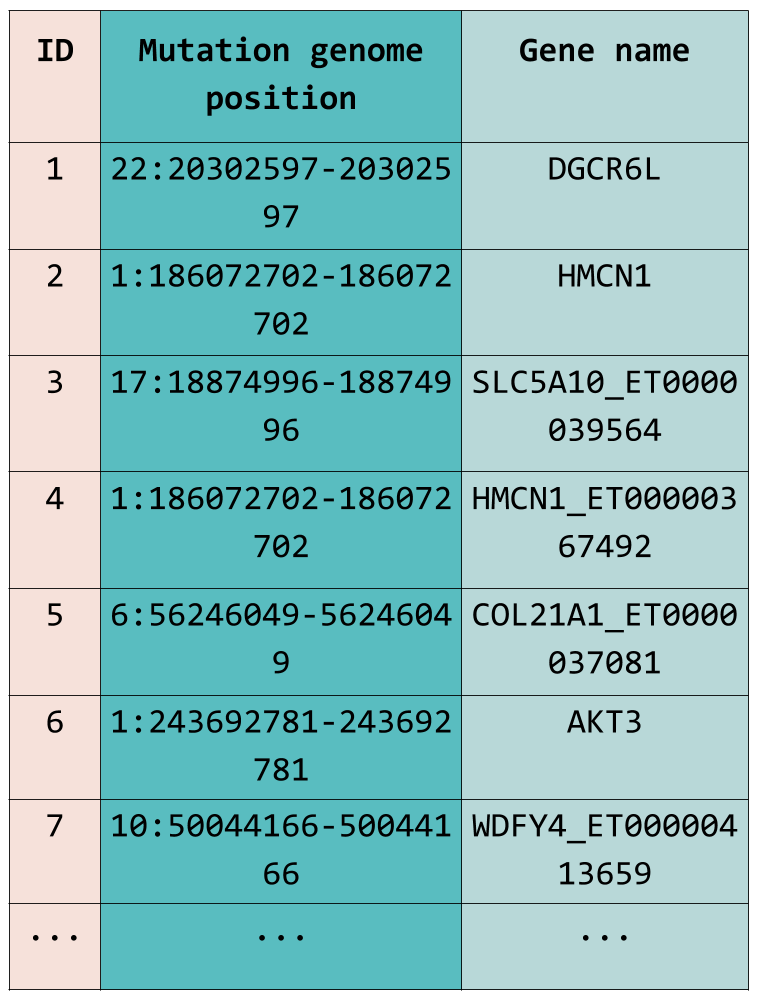}
    \label{fig:projected2}}
  \subfloat[Output1]{
\includegraphics[width=0.32\columnwidth]{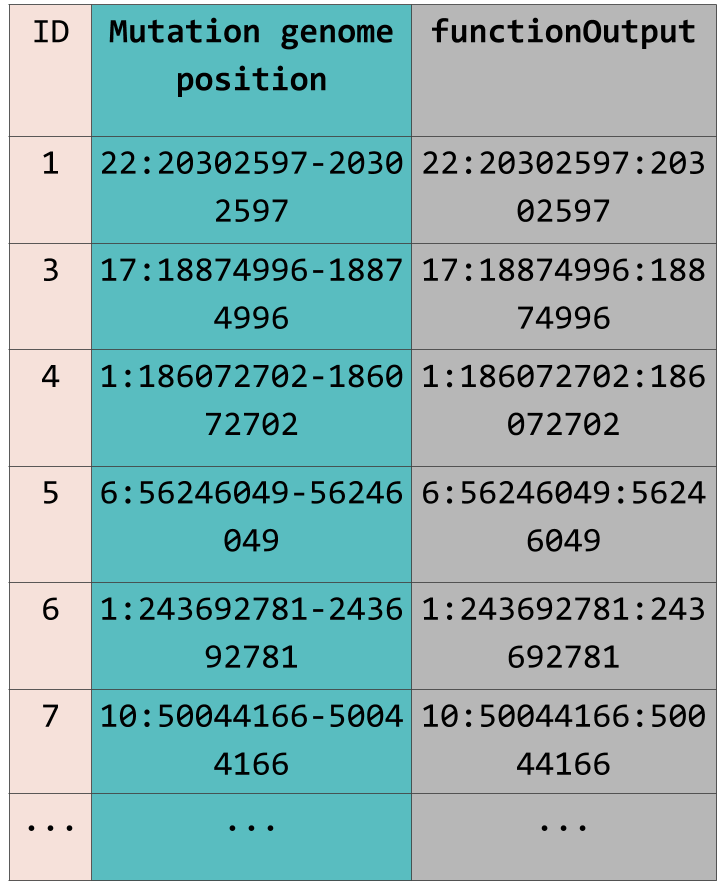}
    \label{fig:output1}}
    \caption{{\bf Transformed sources generated by FunMap.} The DTR2 generates a new source by projecting attributes for each TripleMap (Figures a and b) while DTR1 projects input and output attributes of each FunctionMap into a new source (Figure c). Both remove the generated duplicates.}
    \label{fig:DTR}
\end{figure}

\noindent \textbf{DTR2: Projection of Non-Functional Attributes:}
FunMap provides an additional DTR to further optimize the knowledge graph creation process. Exploiting transformation rules that are proposed in~\cite{JozashooriV19}, FunMap projects all attributes in $S_i$ that are needed by \verb|TriplesMap| including those that are received by \verb|FunctionMap| as input into a new data source $S_i^{project}$ which is added to $S'$. 
To better conceive DTRs, consider the original mappings in \autoref{fig:objectBased} (left-side) and corresponding data source \verb|source1.csv| that can be seen in \autoref{fig:OriginalDatasource}. As shown in \autoref{fig:objectBased}, \verb|FunctionMap1| receives \verb|Mutation genome position| as input. According to DTR1, FunMap projects \verb|Mutation genome position| from \verb|source1| into a new data source named \verb|output1.csv| which is shown in \autoref{fig:DTR}.c. The rows number 2 and 4 have the same value for attribute \verb|Mutation genome posit|-\verb|ion| which leads FunMap to remove the duplicated value from \verb|output1.csv|. Afterwards, \verb|FunctionMap1| is evaluated given \verb|output1.csv| as input and the output values are inserted as a new attribute named \verb|functionOutput| into the \verb|output1.csv| data source. Moreover, attributes \verb|GENOMIC_MUTATION_ID| and \verb|Primary site| from \verb|source1.csv| that are in \verb|TriplesMap1| are projected into the new data source that is shown in \autoref{fig:DTR}.a and duplicated values are removed. Similarly, \verb|Projected2.csv| is created based on the attributes of \verb|TriplesMap2|.

\noindent \textbf{Mapping Transformation Rules (MTRs)}
Mappings are transformed to create the same knowledge graph utilizing the transformed data sources. MTRs are defined considering the role of a transformation function $F_i$ in each \verb|TriplesMap| $T_i$.
\begin{inparaenum}[\bf I\upshape)] 
\item $F_i$ as an \verb|ObjectMap|: 
We refer to the MTRs that are required in this case as \verb|Object-based|. First of all, for each $F_i$, a new \verb|TriplesMap| $T'_i$ is created; it refers to the data source generated as the outcome of $F_i$, i.e., $S_i^{output}$. Accordingly, the \verb|SubjectMap| of $T'_i$ refers to the output attributes $o_i$ in $S_i^{output}$. Afterwards, in \verb|TriplesMap| $T_i$ where $F_i$ is presented as an \verb|ObjectMap|, $F_i$ is replaced by a \verb|joinCondition| which joins $T_i$ and $T'_i$ over attributes $a'_i$, i.e., the input attributes of $F_i$. Moreover, the \verb|logicalSource| of $T_i$ is changed to $S_i
^{project}$, i.e., the corresponding projected data source provided as an outcome of DTR2. 
\item $F_i$ as a \verb|SubjectMap|: Contrary to the \verb|Object-based|, in this set of MTR - we refer to as \verb|Subject-based|- for each \verb|predicateObjectMap| that follows a $F_i$ of the type \verb|SubjectMap|, a new \verb|TriplesMap| $T'_i$ refers to the data source $S_i^{project}$ which is generated as an outcome of DTR2 by projecting the attribute $a'_i$ from $S_i$ that are referenced as \verb|objectMap| in the original \verb|predicateObjectMap|. The \verb|subjectMap| of $T'_k$ --the transformed $T_i$  -- refers to the $o_i$ and its \verb|logicalSource| is $S_i^{output}$.   
Note that \verb|subjectMap| of $T'_i$ is by definition a \verb|TermMap|, which means that its value can be any RDF term according to the RML specification.
\end{inparaenum}
\begin{figure}[t!]
\centering
\includegraphics[width=0.9\textwidth]{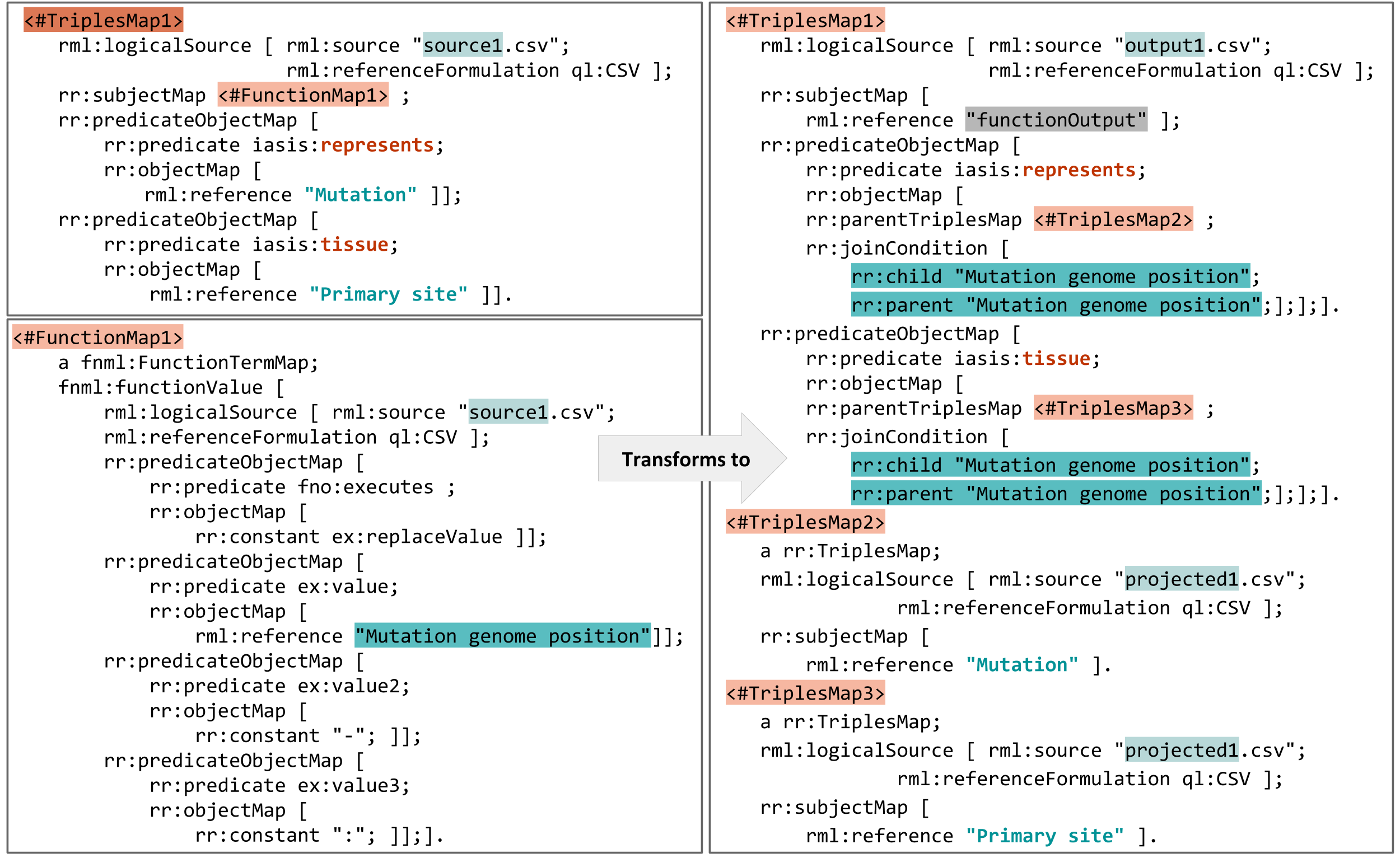}
\caption{\textbf{Example of Subject-based MTR.} An example of mappings including a TriplesMaps and FunctionMap are illustrated on the left and their transformed version including three TriplesMap are shown on the right side.} 
\label{fig:subjectBased}
\end{figure}
Each \verb|objectMap| in $T_i$ that is a \verb|FunctionMap| is replaced by a \verb|joinCondition| between $T_i$ and corresponding $T'_i$ over input attributes $a'_i$ of $F_i$. 
In both cases, the transformed $T_i$-- denoted as $T'_k$-- and $T'_i$ are added to $M'$ and $T_i$ is removed from $M'$. \\     
Figures \ref{fig:objectBased} and \ref{fig:subjectBased} illustrate two examples of rewritten mappings based on DTRs and MTRs. In the left side of both figures, the original mappings are presented while the transformed mappings are depicted on the right side. In the transformed mappings in \autoref{fig:objectBased}, \verb|TriplesMap3| is created for \verb|FunctionMap1|; it refers to the attribute \verb|functionOutput| in the projected data source \verb|output1.csv|-- shown in \autoref{fig:DTR}.c. Then, \verb|FunctionMap1| is replaced in both \verb|TriplesMap1| and \verb|TriplesMap2| by a join condition over the attribute \verb|Mutation genome position| which is the input attribute of \verb|FunctionMap| in the original mapping file as it is highlighted by the same \textcolor{metallicSeaweed}{\textbf{color}}. Accordingly, data sources -\textcolor{powderBlue}{\textbf{highlighted}}- of \verb|TriplesMap|s are also transformed to refer to the projected data sources. 
Consider \autoref{fig:subjectBased} where \verb|FunctionMap| is a \verb|subjectMap|. In both \verb|predicateObjectMa|-\verb|p|s of \verb|TriplesMap1|, \verb|FunctionMap1| is replaced by a \verb|joinCondition| over the attribute \verb|Mutation genome position| that is the input of \verb|FunctionMap1|. To better clarify the performed transformation, consider the first \verb|predicateObjectMap| in \verb|TripleMap1| in the original mappings; the \verb|predicate| is \textcolor{burntSienna}{\textbf{represents}} and the \verb|ObjectMap| refers to the attribute \textcolor{metallicSeaweed}{\textbf{Mutation}}. After the transformation, the first \verb|predicateObjectMap| has the same \verb|predicate| \textcolor{burntSienna}{\textbf{represents}} and through the \verb|joinCondition| refers to the same attribute \textcolor{metallicSeaweed}{\textbf{Mutation}} in \verb|projected1.csv|. 

Pre- and post-conditions of Data Source Transformation Rules (DTRs) and Mapping Transformation Rules (MTRs) are stated in the following properties:
\begin{property} (Lossless Function)
\label{property:p1}
Given data integration systems $DIS_G$=$\langle O,S,M \rangle$ and  $DIS_G'$=$\langle O,S',M \rangle$ such that $DIS_G'$ is the result of applying one DTR1 transformation to $DIS_G$. Then, there are data sources $S_i$ and 
$S_i^{output}$ in $S$ and $S'$, respectively, and the following statements hold:
\begin{itemize}
    \item $S'- S=\{S_i^{output}\}$, there is a mapping $T_i$ in $M$ with a function $F_i$, and \textit{Attrs} contains the attributes $a'_i$ of $F_i$ in $S_i$ and the output attributes $o_i$ of $F_i$. 
    \item  $S_i^{output}$ comprises the attributes \textit{Attrs} and $\pi_{a'_i}(S_i^{output})$=$\pi_{a'_i}(S_i)$.  
    \item For each tuple $t_{i,j}$ in $S_i^{output}$, the values of the attributes $o_i$ in $t_{i,j}$ correspond to the result of $F_i$ over the values of $a'_i$ in $t_{i,j}$, i.e., $t_{i,j}.o_i$=$F_i(t_{i,j}.a'_i)$. 
    
\end{itemize}
\end{property}

\begin{property} (Lossless Projection)
\label{property:p2}
Given data integration systems $DIS_G$=$\langle O,S,M \rangle$ and  $DIS_G'$=$\langle O,S',M \rangle$ such that $DIS_G'$ is the result of applying one DTR2 transformation to $DIS_G$. Then, there are data sources $S_i$ 
and $S_i^{project}$ in $S$ and $S'$, respectively, and the following statements hold:
\begin{itemize}
    \item $S'- S=\{S_i^{project}\}$, and there is a mapping $T_i$ in $M$ defined over the attributes \textit{Attrs} from $S_i$, and $S_i^{project}=\pi_{\textit{Attrs}}(S_i)$. 
\end{itemize}
\end{property}

\begin{property} (Lossless Schema-Ontology Alignments)\footnote{Similarly, this property can be stated for the result of applying MTR over the subject position of a property in a mapping of a data integration system.}
\label{property:p3}
Given data integration systems $DIS_G$=$\langle O,S,M\rangle$ and  $DIS_G'$=$\langle O,S,M'\rangle$ such that $DIS_G'$ is the result of applying one MTR transformation to $DIS_G$. Then, there are \verb|TriplesMap|s $T_i$ in $M$, 
and  $T'_i$ and $T'_k$ in $M'$, and the following statements hold:
\begin{itemize}
    \item $M - M'=\{T_i\}$ and $M'- M=\{T'_i,T'_k\}$. 
    \item There is a function $F_i$ in $T_i$ as the \verb|ObjectMap| of a  \verb|PredicateMap| $p$, and there is a data source $S_i^{output}$ in $S$ which is the \verb|LogicalSource| of $T_i$. The attributes of $S_i^{output}$ are the union of $a'_i$ and $o_i$, while $a'_i$ and $o_i$ are input and output attributes of $F_i$, respectively.
    \item $T_i$ and $T'_k$ are defined over the same \verb|LogicalSource|  $S_i^{project}$. $S_i^{output}$ is the \verb|LogicalSource| of $T'_i$ and $o_i$ is the \verb|SubjectMap| of $T'_i$.  
    \item $T_i$ and $T'_k$ only differ on the \verb|ObjectMap| $p$. In $T_i$, \verb|ObjectMap| of $p$ is defined as $F_i$, while in $T'_k$, a \verb|joinCondition| to $T'_i$ on $a'_i$ defines the \verb|ObjectMap| of $p$. 
\end{itemize}
\end{property}


\section{Experimental Evaluation}
\label{sec:eval}
We evaluate FunMap\footnote{\url{https://doi.org/10.5281/zenodo.3993657}} in comparison to current approaches that create a knowledge graph using the specified data sources and RML+FnO mappings. We aim to answer the following research questions: 
\begin{inparaenum}[\bf {\bf Q}1\upshape)]
\item What is the impact of data duplication rate in the execution time of a knowledge graph creation approach? 
\item What is the impact of different types of complexity over transformation functions during a knowledge graph creation process? 
\item How does the repetition of a same function in different mappings affect the existing RML engines?
\item What is the impact of relational data sources in the knowledge graph creation process?
\end{inparaenum}
All the resources used to perform this evaluation are available in our Github repository\footnote{\url{https://github.com/SDM-TIB/FunMap}}. The experimental configuration is as follows:

\noindent\textbf{Datasets and Mappings.}
To the best of our knowledge, there are no testbeds to evaluate the performance of a knowledge graph construction approach that applies functional mappings. Consequently, following the real-world scenario that initially motivated this research, we create our testbed from the biomedical domain. We generate a baseline dataset by randomly selecting 20,000 records from the coding point mutation dataset in COSMIC\footnote{\url{https://cancer.sanger.ac.uk/cosmic} GRCh37, version90, released August 2019} database. We keep all 39 attributes of the original dataset in the baseline dataset, while only five to seven of them are utilized in mappings. In total, four different mapping files are generated consisting of one \verb|FunctionMap| and four, six, eight, or ten \verb|TriplesMap|s with a \verb|predicateObjectMap| linked to the function. To additionally validate FunMap in case of large-sized data, we create another dataset following the same criteria, with 4,000,000 records and the size of about 1.3GB.

\noindent\textbf{Engines.}
The baselines of our study are three different open source RML-complaint engines that are able to execute RML+FnO mappings and have been extensively utilized in multiple applications and tested by the community: SDM-RDFizer v3.0\cite{IJC20}, RMLMapper\footnote{\url{https://github.com/RMLio/rmlmapper-java}} v4.7, and RocketRML\footnote{\url{https://github.com/semantifyit/RocketRML/}} v1.1. \footnote{We name them SDM-RDFizer**(RML+FnO), RMLMapper**(RML+FnO), and RocketRML**(RML+FnO).}. In order to evaluate the impact of transformation rules, we implement FunMap v1.0 on the top of the aforementioned engines with DTR2 optimization as an optional parameter. We refer to the approach which applies FunMap excluding DTR2 as FunMap$^-$\footnote{We name these combined engines as follows: a) FunMap: FunMap+SDM-RDFizer, FunMap+RMLMapper, and FunMap+RocketRML; b) FunMap$^-$: FunMap$^-$+SDM-RDFizer, FunMap$^-$+RMLMapper, and FunMap$^-$+RocketRML.}. We created a docker image per tested engine for reproducibility.  

\noindent\textbf{Metrics.} \textit{Execution time:} Elapsed time spent by an engine to complete the creation of a knowledge graph and also counts FunMap pre-processing; it is measured as the absolute wall-clock system time as reported by the \verb|time| command of the Linux operating system. Each experiment was executed five times and average is reported. The experiments were executed on an Ubuntu 16.04 machine with Intel(R) Xeon(R) Platinum 8160, CPU 2.10GHz and 700Gb RAM. 

\noindent\textbf{Experimental setups.} Based on our research questions, we set up in overall 198 experiments as the combinations of the following scenarios. We create two datasets from our baseline with 25\% and 75\% duplicates which means in the 25\% duplicate dataset, 25\% and in the 75\% duplicate dataset, 75\% of the records are duplicated. Additionally, two functions with different levels of complexity are created. We describe the complexity level of the functions based on the number of required input attributes and operations to be performed. Accordingly, ``simple'' function is defined to receive one input attribute and perform one operation, while a ``complex'' function receives two input attributes and completes five operations. In total, we create eight mapping files including four, six, eight, and ten \verb|TriplesMap| and one \verb|FunctionMap| to be either ``simple'' or  ``complex''. Additionally, six experiments using 75\% duplicate datasets of 20,000 and 4,000,000 records and a mapping file including ten complex functions are set up in order to be run over a relational database (RDB) implemented in MySQL 8.0 \footnote{https://www.mysql.com/}.   

 \begin{figure}[t!]
 \centering
    \subfloat[SDM-RDFizer - 25\% of duplicates]{
        \includegraphics[width=0.46\columnwidth]{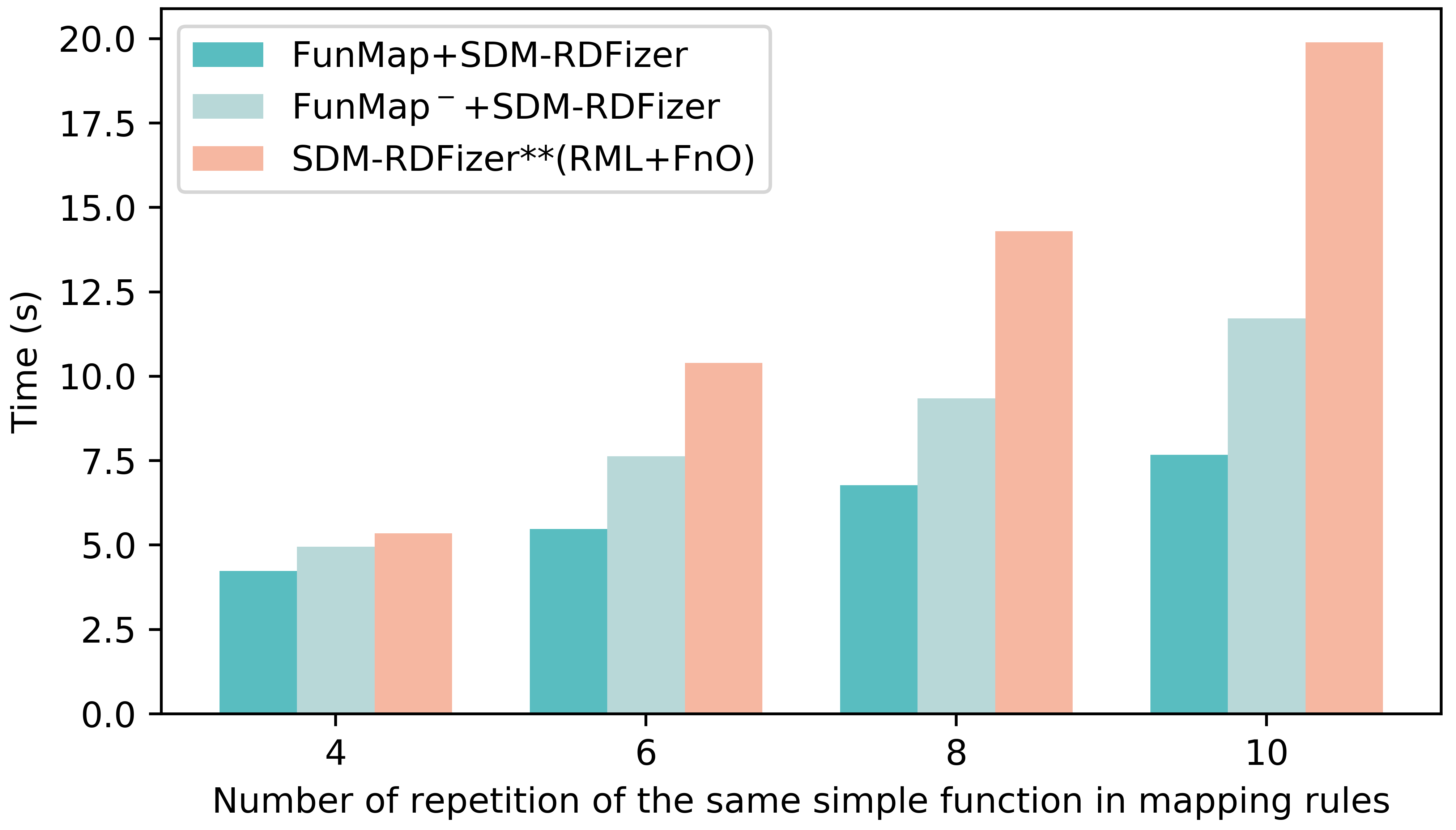}
            \label{fig:vera25_sdmrdfizer}}
    \subfloat[SDM-RDFizer - 75\% of duplicates]{
        \includegraphics[width=0.45\columnwidth]{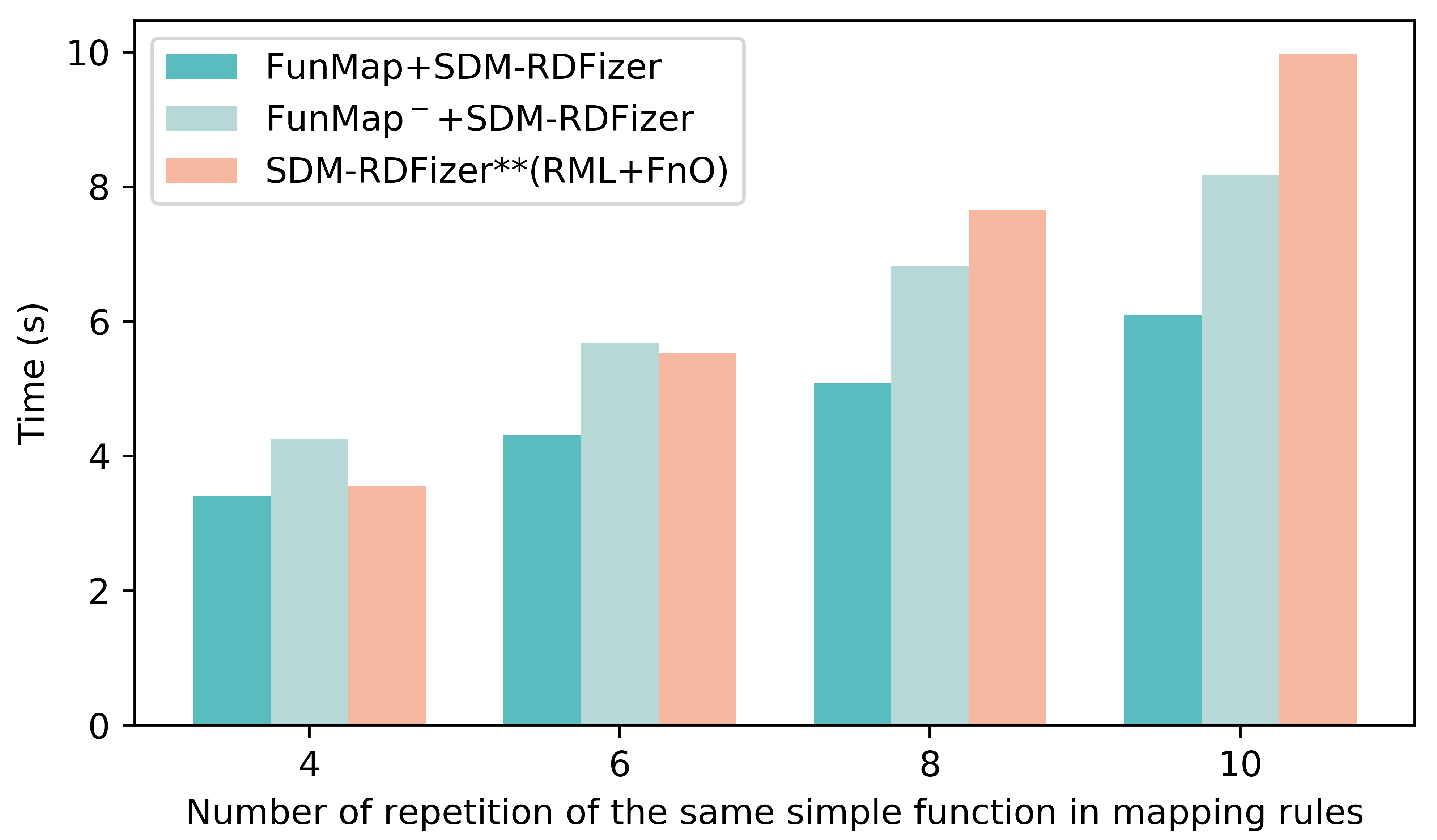}
            \label{fig:vera75_sdmrdfizer}}
\\            
    \subfloat[RMLMapper - 25\% of duplicates]{
        \includegraphics[width=0.45\columnwidth]{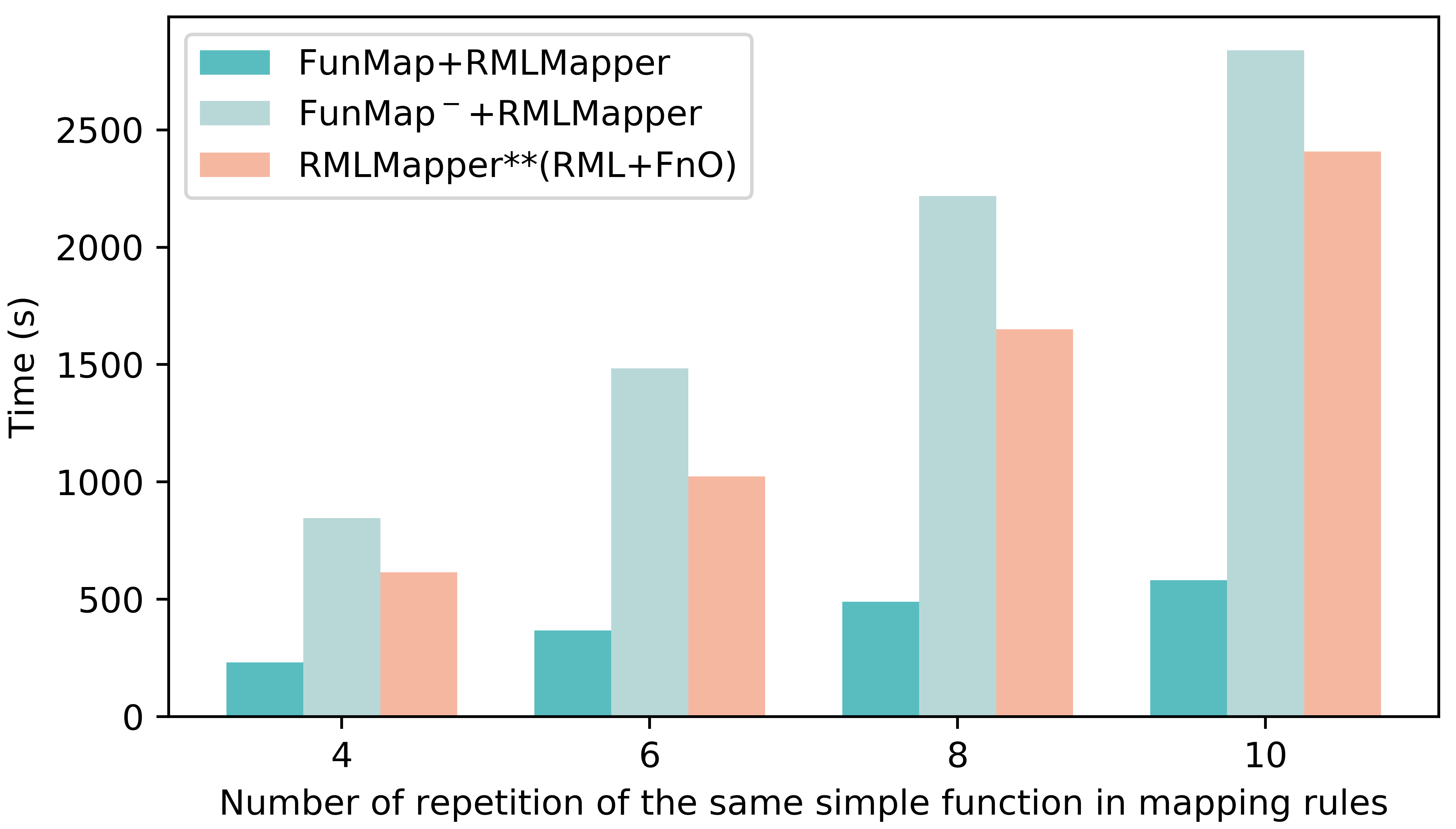}
            \label{fig:vera25_rmlmapper}}  
    \subfloat[RMLMapper - 75\% of duplicates]{
        \includegraphics[width=0.45\columnwidth]{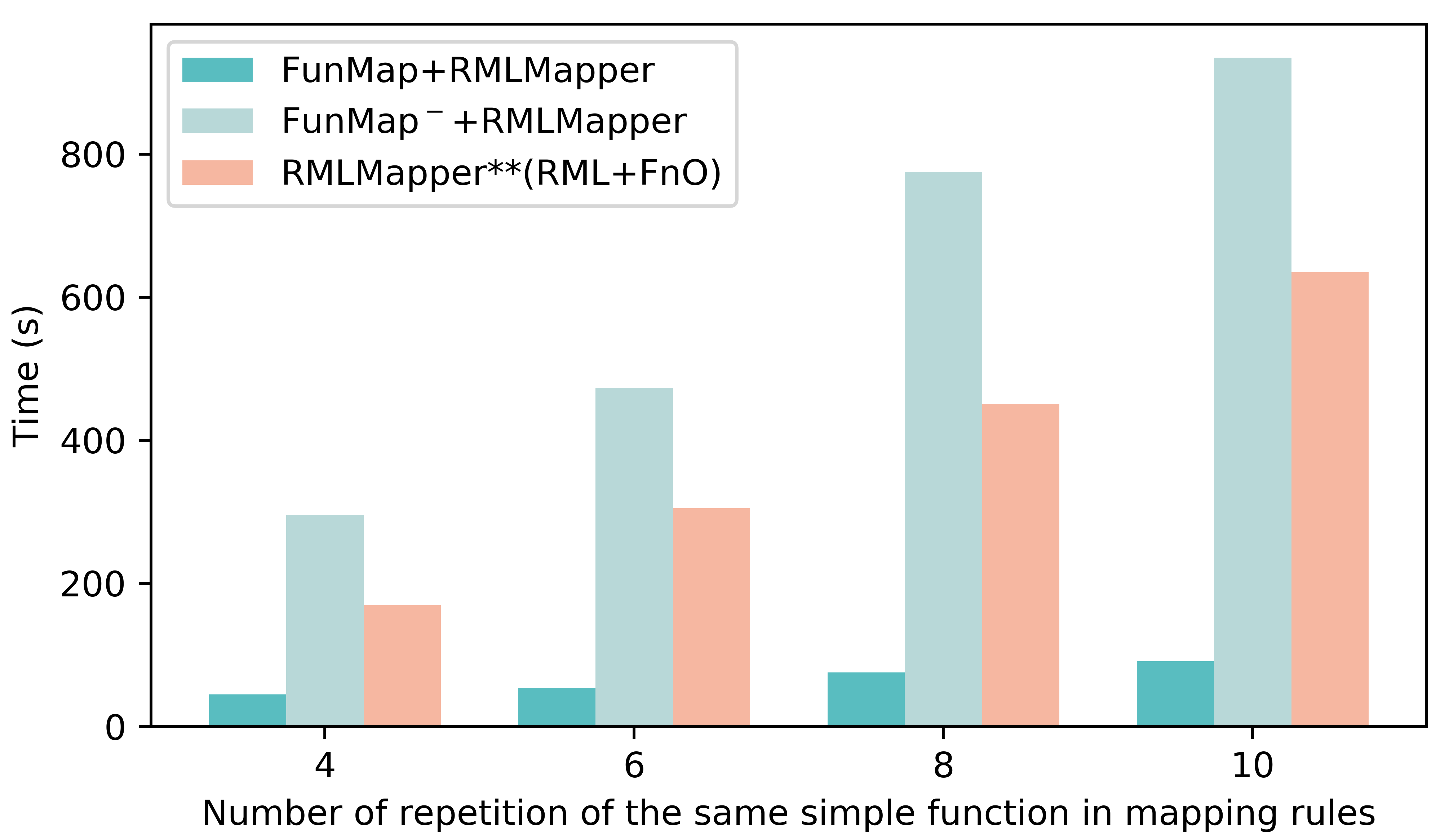}
            \label{fig:vera75_rmlmapper}} 
\\
    \subfloat[RocketRML - 25\% of duplicates]{
        \includegraphics[width=0.45\columnwidth]{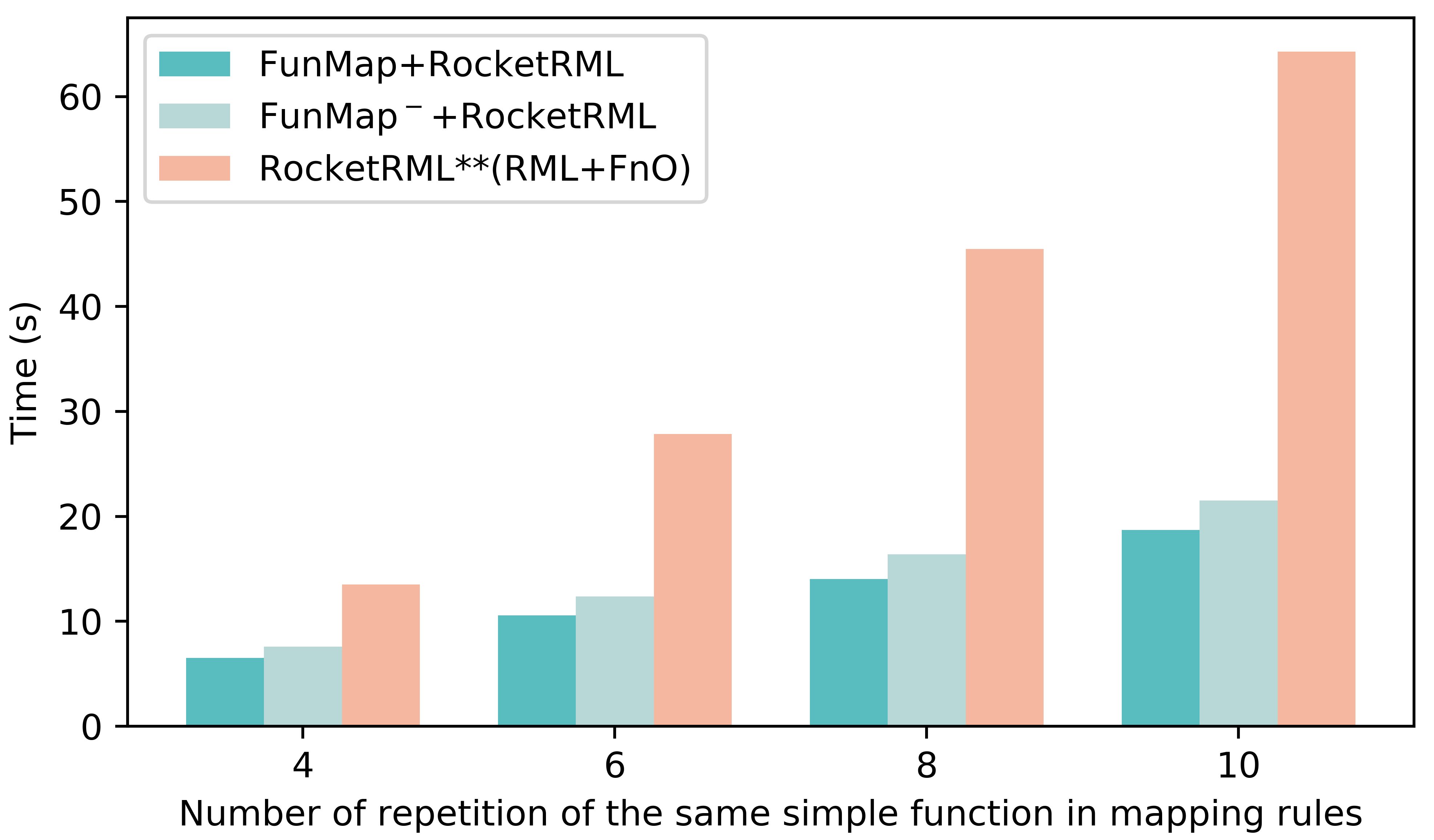}
            \label{fig:vera25_rocketrml}}
    \subfloat[RocketRML - 75\% of duplicates]{
        \includegraphics[width=0.45\columnwidth]{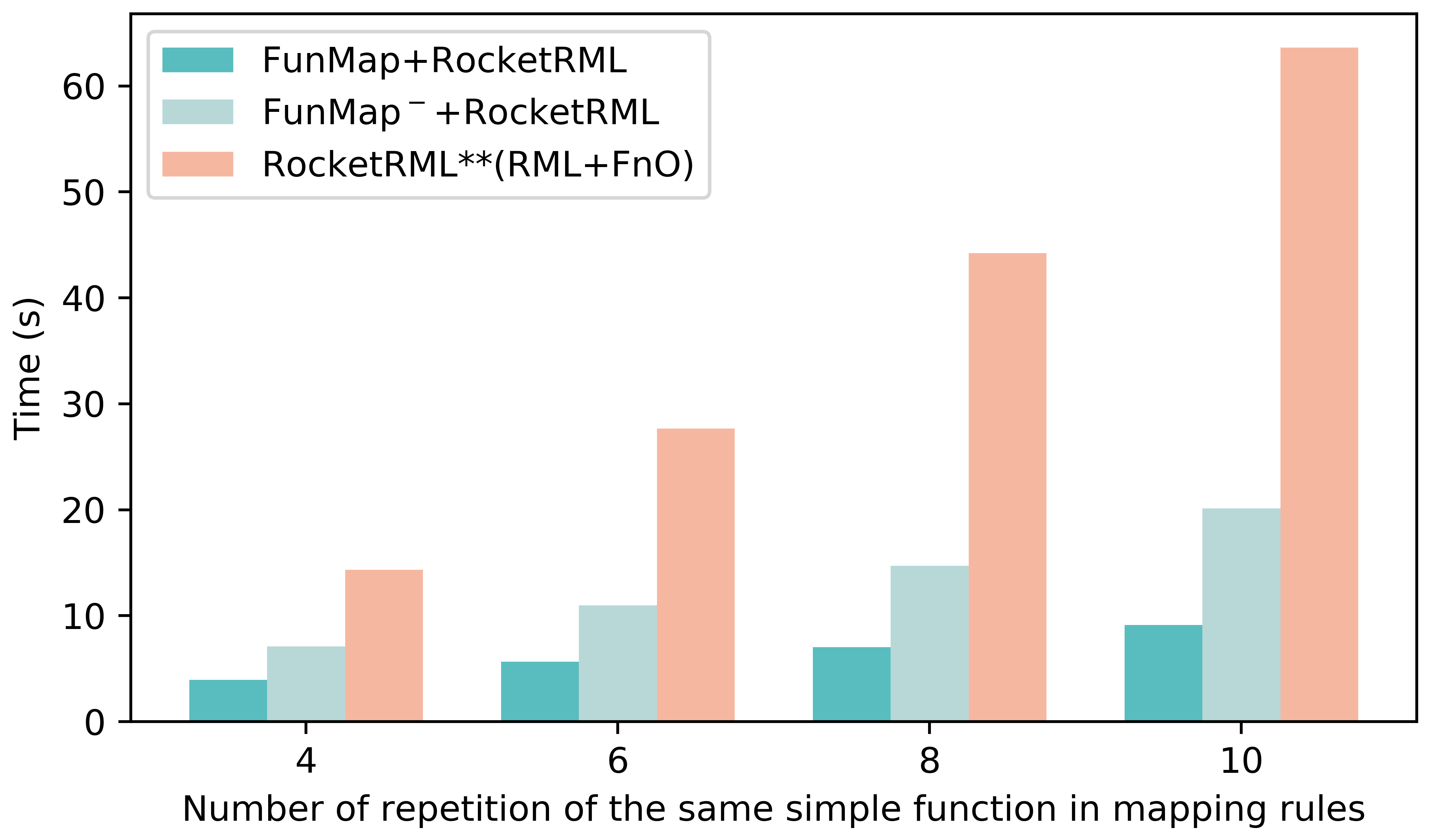}
            \label{fig:vera75_rocketrml}}
\caption{{\bf Total execution time of experiments with simple functions 25-75\% of duplicates.} SDM-RDFizer, RMLMapper and RocketRML executing simple functions in RML+FnO mappings and with FunMap and FunMap$^-$.}
    \label{fig:exp-simple}
\end{figure}

\subsection{Discussion of Observed Results}
\begin{figure}[t!]
 \centering
    \subfloat[SDM-RDFizer - 25\% of duplicates]{
        \includegraphics[width=0.45\columnwidth]{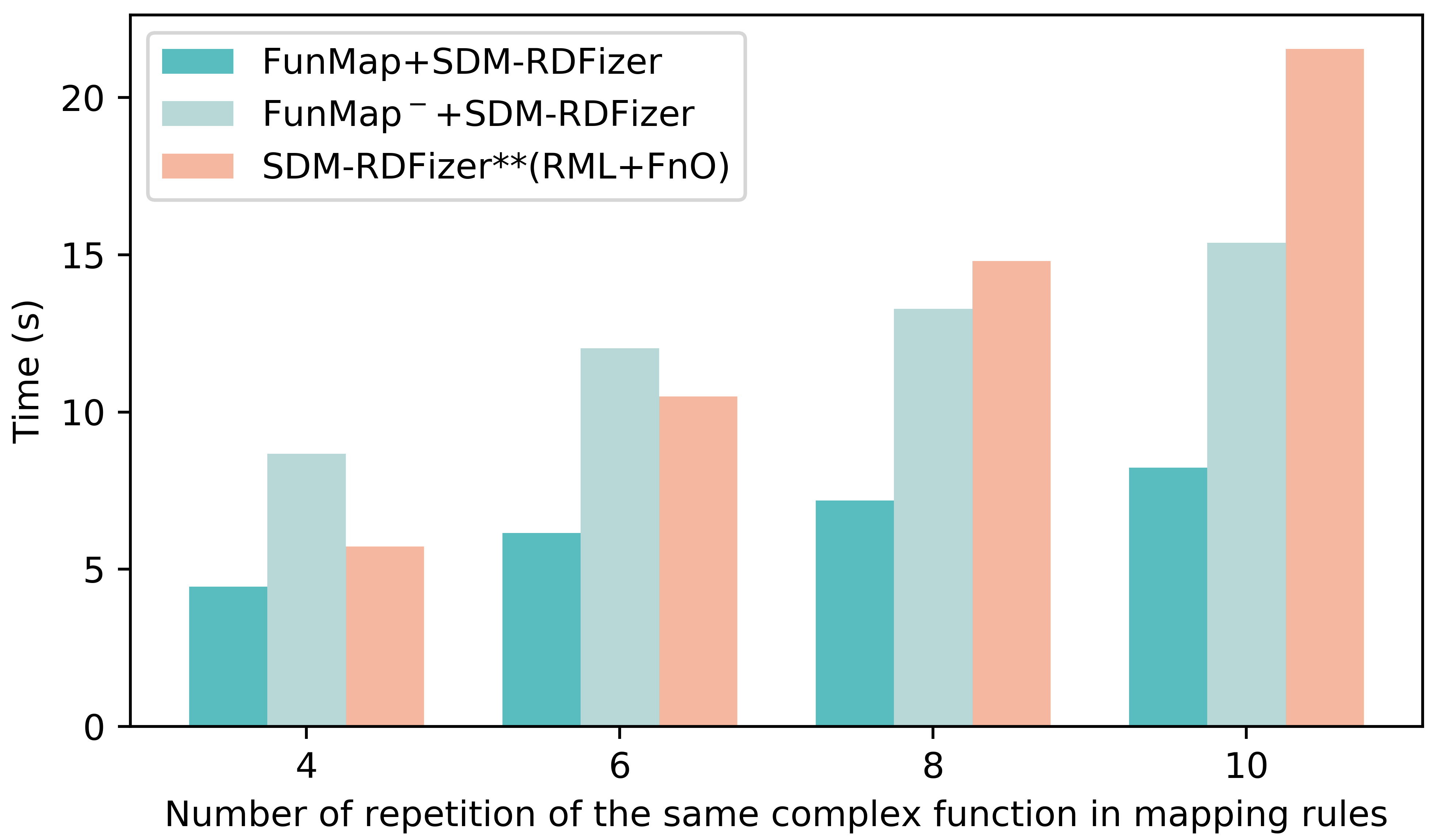}
            \label{fig:vera25_sdmrdfizer_mapsdi}}
    \subfloat[SDM-RDFizer - 75\% of duplicates]{
        \includegraphics[width=0.45\columnwidth]{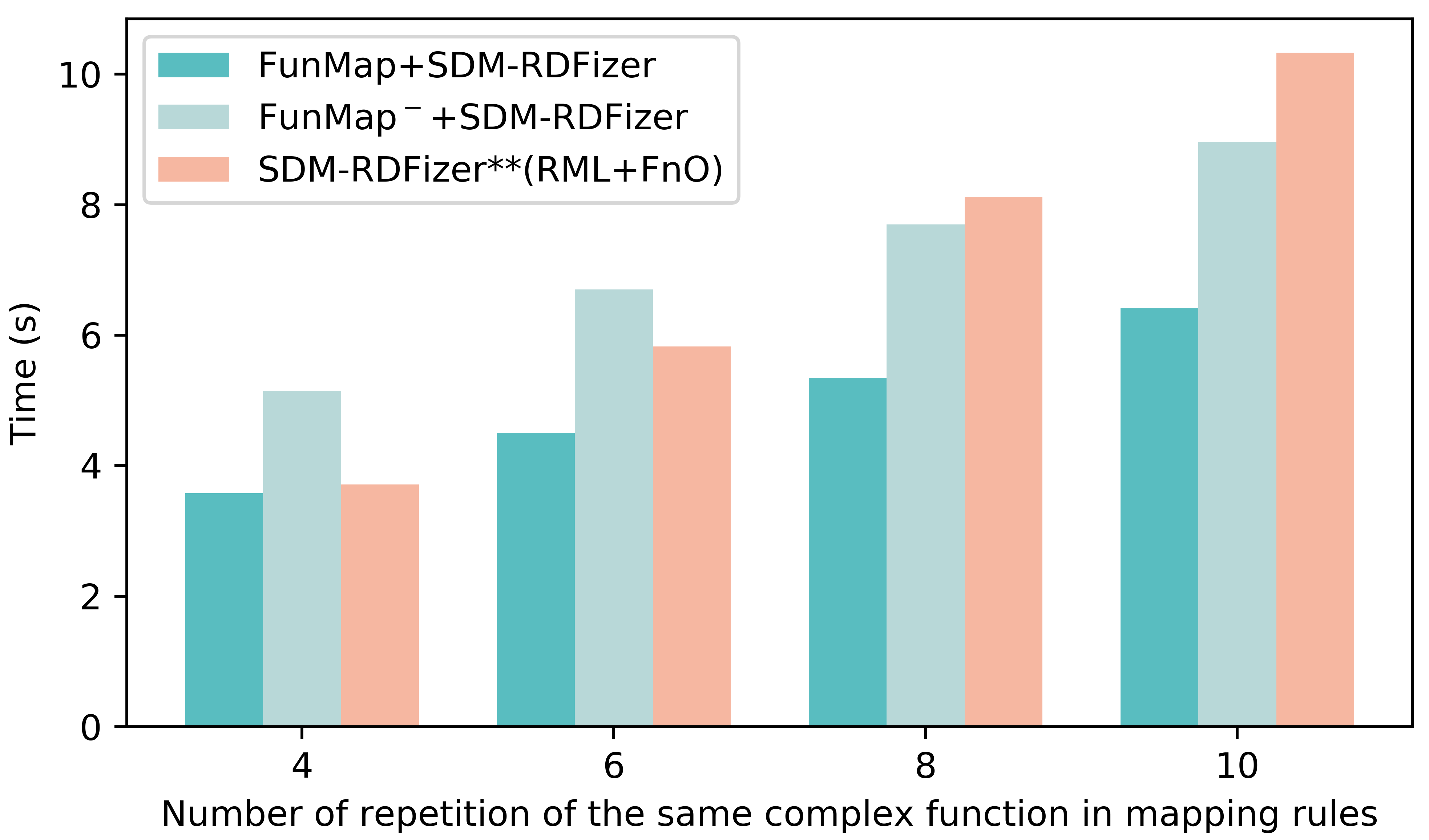}
            \label{fig:vera75_sdmrdfizer_MapSDI}}
\\            
    \subfloat[RMLMapper - 25\% of duplicates]{
        \includegraphics[width=0.45\columnwidth]{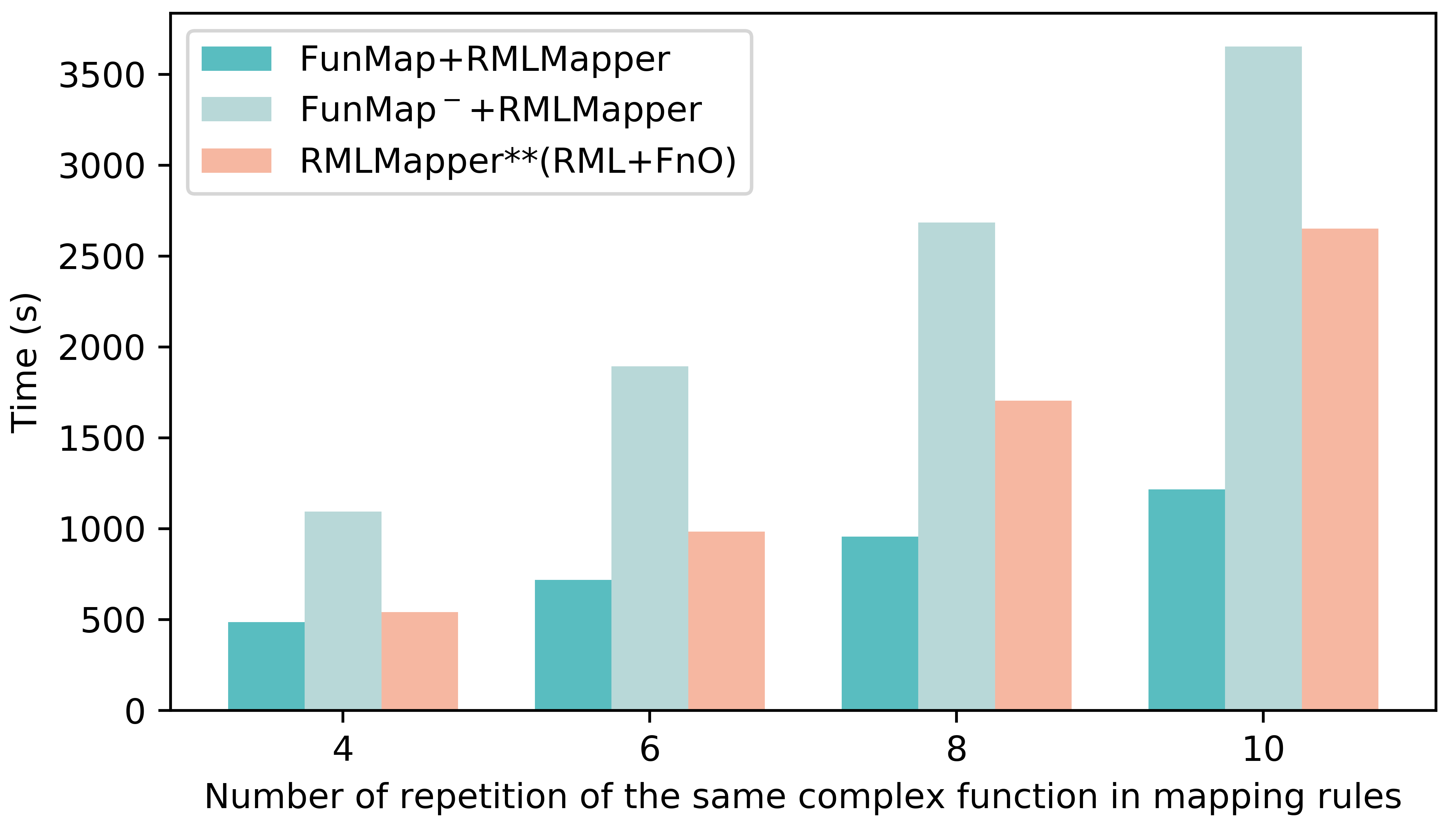}
            \label{fig:vera25_rmlmapper_MapSDI}}  
    \subfloat[RMLMapper - 75\% of duplicates]{
        \includegraphics[width=0.45\columnwidth]{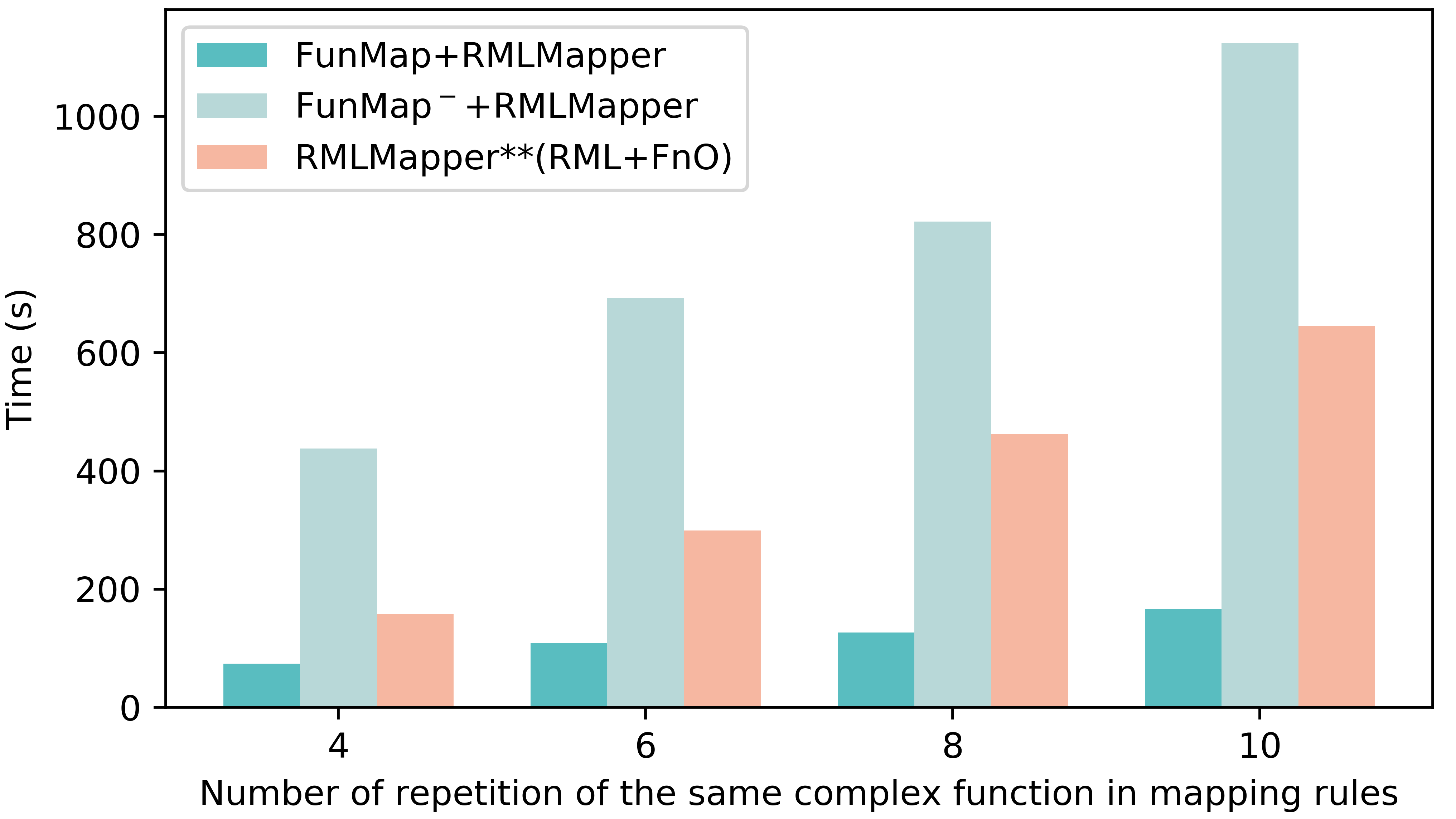}
            \label{fig:vera75_rmlmapper_MapSDI}} 
\caption{{\bf Total execution time for complex functions 25-75\% of duplicates.} SDM-RDFizer and RMLMapper executing complex functions in RML+FnO mappings and with FunMap and FunMap$^-$.}
    \label{fig:exp-complex}
\end{figure}

In this section, we describe the outcomes of our experimental evaluation. Figure \ref{fig:exp-simple} reports on the execution time of the different testbeds in which the functions are considered to be ``simple'' whereas Figure \ref{fig:exp-complex} shows the experiments involving ``complex'' functions. Both figures represent the total execution time for constructing the knowledge graph applying selected engines (i.e., SDM-RDFizer, RMLMapper, and RocketRML) in three different configurations: a) the current version of the engine that is able to directly interpret RML+FnO mappings in the engine (e.g., RMLMapper**(RML+FnO)); b) FunMap$^-$ in conjunction with the engine (e.g., FunMap$^-$+RMLMapper); and c) FunMap together with the engine (e.g., FunMap+RMLMapper). In the case of all the configuration of RocketRML, we only provide the results for the execution of simple functions because the engine does not execute joins with multiple conditions\footnote{Check an example in the zip file of the supplementary material.} correctly, hence, the proposed optimizations cannot be applied. For the rest of the experiments, we have verified that the results are the same for all the approaches in terms of cardinality and correctness. 
The results obtained by the application of SDM-RDFizer with the repetition of simple functions (Figures \ref{fig:vera25_sdmrdfizer} and \ref{fig:vera75_sdmrdfizer}) reflect an improvement of the execution time when FunMap is applied in the process. With the growth of number of duplicates and repeated functions, the difference between the performance of SDM-RDFizer**(RML+FnO) and FunMap+SDM-RDFizer increases. Using this engine, FunMap$^-$ shows the same behavior as FunMap, however, in the case of having a large number of duplicates and a few repeated functions FunMap$^-$ does not improve the performance of SDM-RDFizer**(RML+FnO). 
In the case of using RMLMapper (Figures \ref{fig:vera25_rmlmapper} and \ref{fig:vera75_rmlmapper}), we observe that the results obtained together with FunMap$^-$ (i.e., DTR1 optimization) do not show better performance than RMLMapper**(RML+FnO). DTR1 which only focuses on transforming functions, delegates the removal of the duplicates to the engine which is not accomplished efficiently by RMLMapper. However, in FunMap+RMLMapper, that includes DTR1 and DTR2 optimizations, duplicates are removed before the execution of the RML mappings and leads to obtain the results that clearly show improvements with respect to the baseline. In the same manner as the SDM-RDFizer, the number of repetitions of the functions affects the execution time of the RMLMapper**(RML+FnO), while FunMap maintains similar execution times. Finally, RocketRML (Figures \ref{fig:vera25_rocketrml} and \ref{fig:vera75_rocketrml}) seems not to be affected by the number of duplicates over the input data, obtaining similar execution times for 25\% and 75\% rate for RocketRML**(RML+FnO). However, the number of repetitions over functions impacts the performance of RocketRML**(RML+FnO), increasing the total execution time. The incorporation of DTR1 (i.e., FunMap$^-$+RocketRML) and DTR2 (i.e., FunMap + RocketRML) enhances the performance and scalability during the construction of the knowledge graph, obtaining a similar behavior as the other two tested engines. 

The effect of function complexity over SDM-RDFizer can be observed in Figures \ref{fig:vera25_sdmrdfizer_mapsdi} and \ref{fig:vera75_sdmrdfizer_MapSDI}. Whenever the number of repetitions is low (4-6), the join with multiple conditions affects FunMap$^-$ + SDM-RDFizer, obtaining worse results than SDM-RDFizer**(RML+FnO). However, if repetitions increase (8-10), DTR1 empowers SDM-RDFizer**(RML+FnO) due the reduction of repeated operations during the evaluation of the mappings. Conversely, FunMap+SDM-RDFizer exhibits better results than SDM-RDFizer**(RML+FnO) in all the testbeds. Finally, the behavior of RMLMapper -- when it has to execute complex transformation functions (Figures \ref{fig:vera25_rmlmapper_MapSDI} and \ref{fig:vera75_rmlmapper_MapSDI})-- is affected in terms of execution time for the configuration FunMap$^-$+RMLMapper in comparison to the case of simple functions. As similar as SDM-RDFizer, the join with several conditions is impacting the performance. However, together with data transformation optimizations, FunMap+RMLMapper outperforms the baseline. 

The experimental results on RDBs show even more significant improvement in the performance of both RMLMapper and SDM-RDFizer in the presence of FunMap. In FunMap+RMLMapper, applying \verb|join|s in the SQL queries that define the \verb|logicalSource|s instead of using \verb|joinCondition|s reduces execution time by up to a factor of 18. These results evidence that \verb|joinCondition|s are not efficiently implemented by RMLMapper, and explain why FunMap+RMLMapper is showing less improvement compared to FunMap+SDM-RDFizer in \autoref{fig:exp-complex}. Moreover, FunMap+SDM-RDFizer successfully performs on the large-sized relational dataset of 1.3GB in 5,670.67 seconds, while SDM-RDFizer**(RML+FnO) cannot create the KG and times out after 10,000 seconds.

In overall, we observe that the configurations that interpret RML+FnO mappings directly are affected by the repetition of the functions and the degree of data duplicates, i.e., execution time monotonically increases with number of functions and data duplication degree. In contrast, the incorporation of FunMap to the engines shows less fluctuated behavior when the data duplication rate increases. Additionally, the studied engines handle the repetition of the functions during the construction of the knowledge graph thanks to the pushing down of the execution of the functions directly over the dataset.
In summary, the observed results indicate that the FunMap heuristics improve the performance of data integration systems and generate solutions to the problem of scaled-up knowledge graph construction. The effectiveness of the proposed transformations has been empirically demonstrated on various RML+FnO and RML-compliant engines. However, we observe that there are cases where the application of DTR1 alone is not enough (i.e., FunMap$^-$), being required the applications of all the transformations (i.e., DTRs and MTRs) to provide an effective solution. 
 
\section{Related Work}
\label{sec:relatedwork}
Solutions provided to the problem of KG creation from (semi)-structured data are gaining momentum in practitioners and users~\cite{corcho2020towards}. The seminal paper by Lenzerini~\cite{lenzerini2002data} provides a formal framework for solving this problem and represents a pivot for Ontology-Based Data Access/Integration (OBDA/I)~\cite{PoggiLCGLR08} and the corresponding optimization approaches. Gawriljuk et al. \cite{gawriljuk2016scalable} present a framework for incremental knowledge graph creation. While optimized SPARQL-to-SQL query translation techniques~\cite{chebotko2009semantics} are implemented to support virtual knowledge graph creation~\cite{calvanese2017ontop,PriyatnaCS14}. Albeit efficient, these approaches do not support data transformations (i.e., functions), preventing an efficient evaluation of declarative knowledge graph creation processes. Although FunMap is focused on customized transformation functions for materialized KGs, it can be applied on top of these OBDA approaches. However, this would require the definition of new transformation rules able to push down the corresponding functions into SQL engines.

Rahm and Do~\cite{rahm2000data} have reported the relevance of data transformations expressed with functions during data curation and integration. Grounding on this statement, different approaches have been proposed for facilitating the definition of functions to enhance data curation (e.g., \cite{galhardas2001declarative,GuptaSKGTM12,raman2001potter}). Similarly, declarative languages have been proposed to allow for the definition of functions in the mappings; exemplar approaches include R2RML-F~\cite{debruyne2016r2rml}, FunUL \cite{junior2016funul}, RML+FnO~\cite{de2017declarative}, and D-REPR~\cite{vu2019d}. Moreover, mapping engines enable to interpret functions in declarative mappings (e.g.,  Squerall~\cite{mami2019squerall}, RMLStreamer\footnote{\url{https://github.com/RMLio/RMLStreamer}} and CARML\footnote{\url{https://github.com/carml/carml}} for RML+FnO), as well as in non-declarative formalisms~\cite{lefranccois2016flexible}. FunMap optimizations currently are performed over static data and require the implementation of the \verb|joinCondition| by a KG creation engine. However, RMLStreamer works over streaming data, while CARML does not entirely cover RML joins. Additionally, Squerall is a SPARQL query engine over heterogeneous data able to process RML+FnO on the fly, but Squerall does not implement RML joins. Despite the rich repertory of these approaches, optimizing the declarative description of complex data integration systems remains still open. The absence of frameworks capable of efficiently execute complex data integration systems negatively impacts on the global adoption of existing formalisms in real-world applications of knowledge graphs. FunMap aims at bridging this gap and offering an alternative of evaluating RML+FnO  over existing RML-compliant engines.

\section{Conclusions and Future Work}
\label{sec:conclusions}
We addressed the problem of scaled-up KG creation in complex data integration systems, i.e., systems with large data sources, high data duplication rate, and functional mappings. 
We presented a heuristic-based approach for efficiently evaluating data integration systems with data sources in diverse formats (e.g., CSV or relational). The proposed heuristics are implemented in FunMap, an interpreter of RML+FnO, that converts data integration systems in RML+FnO into equivalent data integration systems specified in RML. Besides shaping an RML-engine independent interpreter of RML+FnO, FunMap generates data integration systems that enhance RML-compliant engines whenever transformation functions are repeatedly used, and data sources are large and have highly-duplicated data. Empirical evaluations of the combination of FunMap with RML-compliant engines suggest that the execution time of RML+FnO can be reduced by up to a factor of 18. Thus, FunMap widens the repertory of tools to scale up knowledge graphs to the enormous increase of incoming data and ease the development of real-world KG applications. As the main limitation, FunMap can only be applied with an RML-compliant engine which supports either \verb|joinCondition| or RDB on the backend. We plan to devise cost-based optimization approaches that, together with the proposed heuristics, allow for the generation of the best solution for a complex data integration system in RML+FnO. 
\noindent\paragraph{\textbf{Acknowledgments.}} This work has been partially supported by the EU H2020 project iASiS No 727658, the ERAMed project P4-LUCAT No 53000015, Ministerio de Economía, Industria y Competitividad, and EU FEDER funds under the DATOS 4.0: RETOS Y SOLUCIONES - UPM Spanish national project (TIN2016-78011-C4-4-R) and by the FPI grant (BES-2017-082511).

\bibliographystyle{abbrv}
\bibliography{biblio}

\end{document}